\documentclass{aa}

\usepackage{graphicx}
\usepackage{txfonts}
\usepackage{lipsum}
\usepackage{subcaption}         
\usepackage{lscape}             
\usepackage{placeins}          

\usepackage{xcolor}

\usepackage{hyperref}

\defcitealias{Salvador-RusinolEtAl2021}{SR21}
\defcitealias{GuerouEtAl2016}{G16}
\defcitealias{BuzzoEtAl2021}{B21}

\begin{document}

    \title{Not so dead after all: young stars in the massive S0 NGC\,3115}

    \author{João P. V. Benedetti\inst{\ref{iac},\ref{ull}}
        \and
        Alexandre Vazdekis\inst{\ref{iac},\ref{ull}}
        \and
        Anna Ferré-Mateu\inst{\ref{iac},\ref{ull}}
        \and
        Elham Eftekhari\inst{\ref{leiden}}
        \and
        Alberto Rodríguez-Ardila\inst{\ref{on}}
        \and
        Rogério Riffel \inst{\ref{ufrgs}}
        }

    \institute{Instituto de Astrofísica de Canarias, Av. Via Lactea s/n, E38205 La Laguna, Spain\\
             \email{joao.pedro.benedetti@iac.es} \label{iac}
            \and Departamento de Astrofísica, Universidad de La Laguna, E-38200 La Laguna, Tenerife, Spain \label{ull}
            \and Leiden Observatory, NL-2300 RA, P.O. Box 9513, Leiden, The Netherlands \label{leiden}
            \and Observatório Nacional, Rua General José Cristino 77, CEP 20921-400, São Cristóvão, Rio de Janeiro, RJ, Brazil \label{on}
            \and Departamento de Astronomia, Universidade Federal do Rio Grande do Sul. Av. Bento Gonçalves 9500, 91501-970 Porto Alegre, RS, Brazil \label{ufrgs}}

    \date{Received September 30, 2025}

    \abstract
    {Massive early-type galaxies (ETGs) host old, metal-rich stellar populations, forming rapidly in the high-$z$ Universe, which has given them the status of red-and-dead. In order to study their properties, however, most works have focused on optical wavelengths. When studied under the ultraviolet (UV) lens, a different picture emerges: a sub-one-per-cent mass fraction of young stars is revealed from detailed spectroscopic analysis.}
    {We aim to measure the mass fraction of young stars in NGC\,3115, a local massive isolated lenticular. Unlike previous UV studies, which focused on massive ellipticals, understanding the spatial distribution of this young component in a lenticular galaxy is a strong test of the ubiquity of this component.}
    {Using SOAR/Goodman spectra down to the atmosphere cut-off, we use line-strength fitting to uncover this hidden population. We use a parametrised star-formation history given by: a single model representing the bulk stellar population, plus a young component modelled by a constant star-formation over the last 1~{Gyr}. Using {\sc ultranest}, we find the posterior distributions for age, metallicity, [Mg/Fe] and young stellar mass fraction.}
    {Using the mismatch in the U-band indices across the different spatial extractions, we measure how the young stellar population mass fraction varies within the galaxy. We find that the gradient peaks at $0.69^{+0.09}_{-0.15}$~{\%} in the centre,  decreasing to zero at around $>1$~{kpc} galactocentric distances. We find that the kinematic and morphological structures correlate with the presence of this young component, dominating in the pressure-supported region. We interpret these findings as residual star formation in the centre of NGC\,3115 likely fuelled by internal processes (e.g. recycled material from past generations of stars). To further strengthen this detection, we also compare models of other UV-bright stellar sources (e.g. metal-rich extreme horizontal branch and metal-poor contributions) that could mimic a young stellar population. We find these models result in worse fits and unphysically high mass fractions to explain the mismatch observed.}
    {The presence of young stars in old, `red-and-dead' galaxies, therefore, seems to be more common than previously expected. The fact that they are being found in a wider population of galaxies and exhibit similar gradients further indicates that residual star formation apparently is a common feature in massive early-type galaxies of any type. Measuring this residual star formation can help place constraints on the physical processes governing the baryon cycle in ETGs.}


    \maketitle
\nolinenumbers

\section{Introduction}

    Early-type galaxies (ETGs) dominate the high-mass end of the local galaxy population (M$_*\gtrsim$10$^{11}$M$_\odot$) and contain more than half of the local stellar mass \citep[][]{FukugitaEtAl1998, KelvinEtAl2014}. Observationally, the characterisation of this population is rather well established, hosting old, metal-rich, and [Mg/Fe]-enhanced stellar populations \citep[][]{GallazziEtAl2005, GallazziEtAl2006, McDermidEtAl2015, Martin-NavarroEtAl2019, Diaz-GarciaEtAl2019}. Their evolution is rooted in a two-phase formation scenario, whereby an early and quick burst of star formation (SF) creates massive compact galaxies in the early Universe. Previously thought to occur at $z>2$, JWST has now shown that these compact massive galaxies are actually more numerous and form faster and earlier than expected, already existing at $z\sim5$ \citep[][]{CarnallEtAl2023, GlazebrookEtAl2024, CarnallEtAl2024}. This dissipative phase is then followed by subsequent accretion of smaller satellites, mainly building their halos and increasing their sizes \citep[][]{TrujilloEtAl2007, BuitragoEtAl2008, OserEtAl2010}. In this scenario, after the intense initial SF burst, which formed the bulk of the stellar population, ETGs are thought to evolve rather passively \citep[e.g.][]{VazdekisEtAl1997}, giving them the status of `red-and-dead' galaxies. 
    
    Despite the large amount of information gathered about the nature of ETGs over decades, whether this is the true and unique evolution these old giants follow is still an unsolved puzzle in modern astrophysics. Cosmological simulations allow us to connect high-$z$ observations to present-day galaxies, as well as providing evolution pathways \citep[e.g.][]{SpringelEtAl2005, SchayeEtAl2015}. Despite their successes in reproducing large-scale cosmological relations, modelling the complex baryonic physics is a much harder task, typically requiring sub-grid physics implementations. Due to the hierarchical nature of halo assembly in simulations, more massive halos take longer to form. Therefore, higher fractions of young stars are predicted for more massive halos \citep[e.g.][]{ThomasEtAl2005, TrayfordEtAl2015}. This is in tension with the downsizing scenario established in observational studies, where the most massive galaxies actually are the fastest to form \citep[][]{CowieEtAl1996, TragerEtAl2000, GallazziEtAl2006}. Active galactic nuclei feedback has been proposed as one of the solutions to this conundrum, as it can inject energy into the surrounding gas, halting SF \citep[][however see \citealt{ChoiEtAl2024}]{CrotonEtAl2006, DeLuciaEtAl2006, BowerEtAl2006}.
    
    The main issue is that constraining any presence of young stars in these old galaxies is quite challenging. Most studies have focused so far on obtaining information from the optical part of the spectrum ($\sim0.4-1.0$~{$\mu m$}), where the bulk of the old stellar population dominates \citep[][]{Renzini2006}. Going towards the ultraviolet (UV), the iron opacity increases in the cool atmospheres of the old metal-rich stars, significantly decreasing the UV flux, which allows for the detection of young stars \citep[][]{VazdekisEtAl2016}. Therefore, harnessing the constraining power from the UV is critical to reveal any hidden contributions from young stars.

    GALEX was a groundbreaking facility allowing for the first time to perform the first space UV sky survey \citep[][]{MartinEtAl2005}, tracing ETGs' UV properties across different parameters. GALEX-based studies started to point to the presence of residual SF in ETGs \citep[][]{YiEtAl2005}, with as much as $\sim 20$~{\%} of their stellar masses possibly having formed at $z<2$ \citep[][]{KavirajEtAl2007a, Kaviraj2008}. While there is evidence for SF in ETGs \citep[e.g.][]{CombesEtAl2007, GomesEtAl2016, BenedettiEtAl2023}, \citet{Salvador-RusinolEtAl2020} confirmed that massive ETGs at $z\sim0.38$ host sub-one-per-cent ($<1\%$) contributions from young stars ($<1$~{Gyr}). Therefore, galaxies that were thought to be completely devoid of any star formation were in fact still forming new stars, albeit at very low, almost negligible, levels. This unexpected result was possible due to the use of SDSS-BOSS data, which provides access to the rest-frame near-ultraviolet (NUV) spectra of ETGs, coupled with the extension towards the UV of the E-MILES models \citep[][]{VazdekisEtAl2016}, reaching an uncharted regime. \citet[][hereafter \citetalias{Salvador-RusinolEtAl2021}]{Salvador-RusinolEtAl2021} found similar fractions of young components when analysing X-shooter spectra for a stack of six Brightest Cluster Galaxies (BCGs). They also showed that the young component seems to be concentrated towards the centre of the stack. Moreover, \citet{Salvador-RusinolEtAl2022} analysed the U-band features of NGC\,1277, a prototypical relic galaxy. By definition, this is a galaxy that avoided the accretion/merger-driven second phase described above. As no SF is expected to occur after $z\sim$2 for this relic galaxy \citep[][]{TrujilloEtAl2014, Ferre-MateuEtAl2017}, the finding that even in such an extreme object $\sim 0.8$~{\%} of young star mass fraction was found was a surprising result. 
    
    Collectively, these results show that the presence of young stars in old systems is much more common than previously thought. While the ubiquitous presence of young stars in ETGs points towards a common mechanism, the origin behind it is still elusive. The fact that even NGC\,1277 hosts this component suggests that an in-situ process is actually the most likely driver, as recycled material from past stellar generations could re-collapse, forming new stars \citep[][]{DopitaEtAl2000, Salvador-RusinolEtAl2022, SpinielloEtAl2025}.
   
    Nonetheless, exploiting the information retained in the UV of ETGs is challenging, as other UV-bright components could also boost the UV emission, mimicking the presence of young stars. The most notable manifestation from such components is the unexpected UV rise at $\lambda < 2500$~{\AA}, known as the UV-upturn \citep[][]{CodeWelch1979, OConnell1999}. Many stellar sources have been proposed to explain this emission: He-enriched stars, low-metallicity blue horizontal branch (HB) stars, metal-rich Extreme HB (EHB) stars experiencing strong mass loss, post-AGB stars, binaries, etc. \citep[][]{GreggioRenzini1990, MarastonThomas2000, YiEtAl1997, BrownEtAl1997, HanEtAl2006}. Therefore, a careful analysis of the presence of residual SF in ETGs also needs to take into account the possibility of other UV-bright components, engaging in panchromatic studies to differentiate among the possible sources \citep[][]{BonattoEtAl1996, JeongEtAl2009, Salvador-RusinolEtAl2020}. For example, near-infrared (NIR) studies of massive ETGs have shown that strong CO absorption features cannot be straightforwardly interpreted as evidence for intermediate-age populations, since similar CO strengths are also found in the relic galaxy NGC\,1277, where such populations are absent or negligible \citep[][]{EftekhariEtAl2022}. This reinforces the need for panchromatic constraints when interpreting small residual star-forming components in old galaxies.
    
    In order to expand our knowledge on the nature of young stars in ETGs, in this work, we derive, for the first time in an individual galaxy, the spatial behaviour of this young component by means of U-band spectra from SOAR/Goodman. We focus on NGC\,3115, an isolated lenticular galaxy, which is one of the most well-studied massive ETGs in the local Universe. Crucially, this galaxy does not host a UV upturn \citep[FUV-NUV = $1.68 \pm 0.02$][]{BonattoEtAl1996}, unlike previous studies that either did not have the UV classification for their objects \citetalias{Salvador-RusinolEtAl2021} or showed a mild (although not weak) UV upturn \citep[FUV-NUV=$0.75\pm0.11$,][]{Salvador-RusinolEtAl2022}. The intrinsic differences between this galaxy and previously studied ones, both in terms of environment and morphology, make NGC\,3115 the perfect individual benchmark. The detection, if any, of young stars in NGC\,3115 and the study of possible gradients within the galaxy would thus support the universality behind this component and shed light on the evolution of massive ETGs, up to their `last breath'. 
    
    The layout of the paper is as follows. In Sect.~{\ref{sec:ngc3115}} we provide an overview of the studied galaxy, NGC~3115, as well as how observations and data reduction were performed. Sect.~\ref{sec:stelkin} describes the kinematical modelling and how the radial apertures were built to increase the signal-to-noise ratio (S/N) in the U-band spectra, while Sect.~\ref{sec:stelpops} presents the line-strength fitting method to retrieve stellar population properties. In Sect.~\ref{sec:results}, our main findings regarding the presence of young stars in NGC\,3115 are shown, and then discussed by putting them in the broader context and distinguishing them from other UV-bright components in Sect.~\ref{sec:discussion}. Lastly, a summary and the main conclusions are presented in Sect.~\ref{sec:conclusions}.

\section{A test case: NGC\,3115, the closest isolated massive lenticular}

\label{sec:ngc3115}

    NGC\,3115 is the closest field lenticular (S0) \citep[i.e. no other group member with at least one-third of its luminosity;][]{BrodieEtAl2014} and, given its proximity \citep[D\,=\,9.68\,Mpc,][]{TonryEtAl2001}, it has been largely studied at various wavelengths besides the optical. It is one of the most massive S0s known \citep[M$_*\sim$1.1$\times$10$^{11}$M$_{\odot}$,][hereafter \citetalias{GuerouEtAl2016}]{GuerouEtAl2016} and one of the brightest galaxies in the southern hemisphere \citep[B\,=\,9.87\,$mag$,][]{deVaucouleursEtAl1991}. This is especially important as ETGs are particularly faint towards the U-band (between 3000 and 4000~{\AA}). By focusing on a bright galaxy, getting high-quality U-band data is more time-efficient.
    
    The galaxy has a clear three-component structure noted from both photometry and integral-field spectroscopy (IFS) studies \citep[][hereafter \citetalias{BuzzoEtAl2021}]{ArnoldEtAl2011, DolfiEtAl2020, BuzzoEtAl2021}: a central dispersion-dominated spheroidal bulge, a kinematically cold fast-rotating thin disc, and an outer halo. The globular cluster (GC) population in NGC\,3115 was one of the first to be attributed to the two-phase formation scenario due to its distinct red and blue GC populations \citep[][]{ArnoldEtAl2011}, in line with the kinematics of these two populations tracing the in-situ (bulge and the kinematical thin disc) vs. ex-situ (halo) kinematics (\citetalias{GuerouEtAl2016}; \citealt[][]{DolfiEtAl2020}).
    
    The stellar populations also trace these differences \citepalias[][]{GuerouEtAl2016}. For instance, the bulge and thin disc trace metal-richer stellar populations ($+$0.2~{dex} in the innermost regions) and show a clear negative gradient along the major axis of the galaxy. On the contrary, the halo is characterised by ubiquitous metal-poor ($-$0.4~{dex}) stellar populations. The age distribution is distinct: while the bulge and halo exhibit 13~{Gyr}-old populations, the thin disc has slightly younger ages ($\sim$10~{Gyr}), representing a longer timescale for its evolution. This is a key indicator that this galaxy could not have experienced any major mergers in the past $\sim$$10$~{Gyr}, as it would have destroyed the disc configuration and/or severely changed the ages found (\citealt[][]{ArnoldEtAl2011},\citetalias{GuerouEtAl2016},\citealt[][]{DolfiEtAl2020}). While the exact age differences among the different components are small and vary slightly among different studies \citepalias{GuerouEtAl2016, BuzzoEtAl2021}, the fact is that NGC\,3115 is a galaxy dominated by old stellar populations.

    Interestingly, \citetalias{GuerouEtAl2016} found that the thin disc was slightly younger compared to the bulge, while \citetalias{BuzzoEtAl2021} found that the bulge had a bluer NUV-$r$ colour. This gradient was related to an SF episode triggered by a recent ($\sim$1~{Myr}) flyby of a companion galaxy. This would explain the blue colours, the spiral structures seen in the disc, and the displacement in the isophote centres. \citet{PociEtAl2019} suggested that the spiral structures could be the fossil structures from a precursor spiral galaxy that merged $>10$~{Gyr} ago, leaving such an imprint. The presence of an end-on 
    bar has been discussed too, which could also lead to the formation of the spiral structures due to the interactions with its gravitational potential (\citealt[][]{EmsellemEtAl1999}, \citetalias[][]{GuerouEtAl2016}, \citealt[][]{PociEtAl2019}).

    NGC\,3115 hosts the nearest billion-solar-mass black hole defined by a low-luminosity AGN \citep[][]{WrobelNyland2012, AlmeidaEtAl2018}, which has been detected using GMOS observations. Looking at the [NII]+H$\alpha$ complex, it showed faint ionised gas emission lines \citep{MenezesEtAl2014a}. The galaxy has little molecular and atomic gas (1.86$\times$10$^{6}$~{M$_\odot$} and 1.14$\times$10$^{7}$~{M$_\odot$}, respectively; \citealt{WelchSage2003, SageWelch2006}), lower than the expected returned material from previous generations of stars. This galaxy also does not host a substantial hot-gas halo \citep[][]{LiEtAl2011}.

    Lastly, this galaxy has been observed with past UV missions (e.g. HUT, IUE, GALEX). It is known for being UV weak \citep[FUV-NUV = $1.68 \pm 0.02$][]{BonattoEtAl1996} and not hosting a UV upturn according to \citet[][]{YiEtAl2011} criteria.
    
    All in all, NGC\,3115 is a unique galaxy to study the presence of young stars in old galaxies. The lack of a UV upturn discards the presence of exotic UV-bright components in redder wavelengths, and the NUV-$r$ gradient already hints at the presence of young stars. Moreover, its brightness and proximity allow us to get high-quality U-band spectra while preserving the spatial information to derive gradients. Being an isolated galaxy further positions it in sharp contrast to the cluster BCGs of \citet{Salvador-RusinolEtAl2021}, providing a strong test of the presence of young stars in old galaxies.
       
    \subsection*{Observations and data reduction}
    
    NGC\,3115 was observed using the 4.1-m Southern Astrophysical Research (SOAR) Telescope based on Cerro Pachón in the Chilean Andes under the proposal program ID SO2019-013 (PI: Rodríguez Ardila) on 2019 March 10th. The Goodman spectrograph was used to get a 3.9~{arcmins} long-slit spectrum with the 1.0" slit-width, with the blue camera, and the 600 l/mm grating. This resulted in a spectrum covering the $3010-5690$~{\AA} spectral range with a resolving power $\sim 1200$, aligned with the major axis of the galaxy (PA$=45^{\circ}$). The total exposure time was two hours, split into four exposures obtained with a seeing varying between 1.2 and 1.8~{arcsec}.

    {\sc Pypeit} \citep[][]{ProchaskaEtAl2020, Prochaska2025} was used to perform standard reduction procedures, including bias subtraction, cosmic ray removal, flat-fielding, wavelength calibration, sky subtraction and spectral extraction. Wavelength calibration was particularly challenging due to the internal reflections, but {\sc Pypeit} is able to iteratively discard bad lines in the arc files when doing the 2-D wavelength solution, removing any problematic reflections. Using the HgAr lamp, the final wavelength solution had a root-mean square of 0.319~{\AA} ($\sim 19$~{km s$^{-1}$} at 5000~{\AA}). Relative flux calibration was performed using the GD-108 standard star \citep[][]{Oke1990}. An asset from line-strength fitting is the narrow wavelength windows (<100~{\AA}). As instrumental responses tend to be nearly constant in these regimes and with the further inclusion of bracketing pseudo-continua band-passes in index definitions, the method is not heavily dependent on flux calibration.

\section{Stellar kinematics and radial apertures}
\label{sec:stelkin}
    
    After the data were fully reduced, manual extractions were performed along the spatial direction, with apertures as wide as the seeing until reaching an S/N of 10~{\AA$^{-1}$} in the 4800-5500~{\AA} window. With such S/N, it is possible to derive reliable kinematical information from the prominent absorption lines. By modelling the kinematics in the 4800-5500~{\AA} region, we then added several extractions to get a higher S/N in the U-band, particularly in the outer parts. This procedure is especially important in this regime because NGC\,3115 is faint at the bluest wavelengths, while both grism efficiency and atmospheric transmission are lower.

    NGC\,3115 exhibits a well-defined rotation pattern between $\pm 300$~{km\,s$^{-1}$} and its stellar velocity dispersion varies significantly from 300 to 100~{km\,s$^{-1}$} within 2 effective radii \citepalias[R\textsubscript{e}, e.g.][]{GuerouEtAl2016}. This means that to co-add the spectra, one first needs to de-redshift and bring all spectra to the same resolution before the stacking. The full-spectral fitting code {\sc pPXF} was thus used to obtain the stellar kinematic parameters \citep[][]{CappellariEmsellem2004, Cappellari2017}. The code uses an SSP base to fit a given spectrum by optimising the line-of-sight velocity distribution (LOSVD), parametrised by radial velocity (v\textsubscript{r}), velocity dispersion ($\sigma_*$), and the higher Gauss-Hermite moments (e.g. h$_3$ and h$_4$). Here we used the E-MILES models \citep[][]{VazdekisEtAl2016} to fit all the spectra in the 4800-5500~{\AA} region. We broaden the SSP models to match the resolution of the spectrograph (FWHM 4.1~{\AA}) so that the final derived velocity dispersion corresponds only to the kinematical broadening. We employed a 7th-order additive polynomial for the fitting, the lowest possible for which the kinematical parameters stabilised. The same polynomial degree was used for all exactions in order not to introduce any biases.
    
\begin{figure*}[h]
       \centering
       \includegraphics[width=1.0\textwidth]{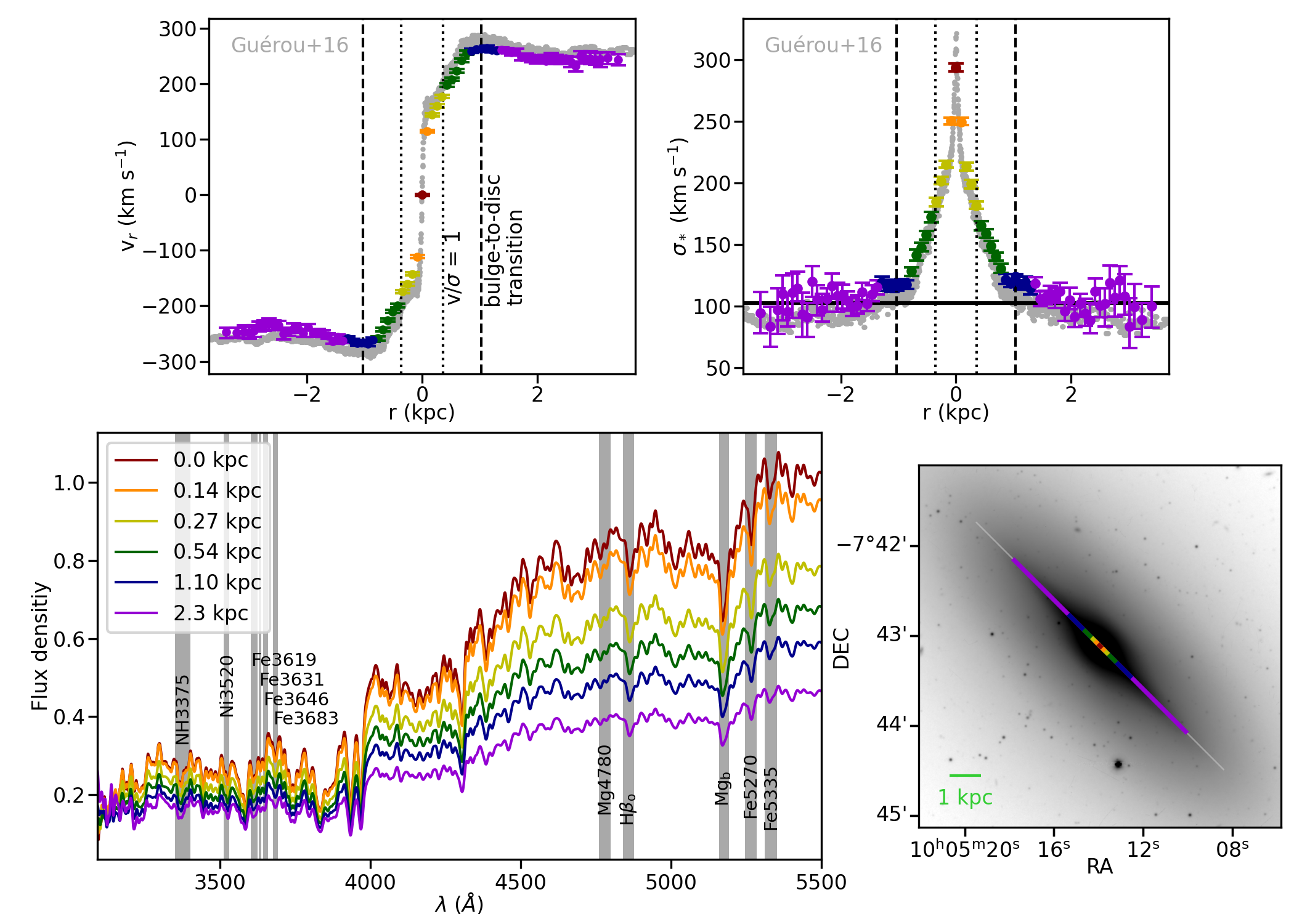}
       \caption{\textbf{Top:} Stellar kinematic properties derived from {\sc pPXF} fitting using a Gaussian LOSVD: radial velocity (\textit{left panel}) and velocity dispersion (\textit{right panel}). Dotted vertical lines indicate the kinematic transition (v/$\sigma$=1) and dashed ones the photometric transition (bulge-to-disc transition from \citetalias{BuzzoEtAl2021}). The horizontal line indicates the spectral resolution of the spectrograph ($\sim100$~{km s$^{-1}$} at 5000~{\AA}). Colours indicate seeing-limited extractions that were co-added for the stellar population analysis. Grey points correspond to the kinematic results from \citetalias{GuerouEtAl2016}. \textbf{Bottom:} \textit{Left:} Resulting annular spectra obtained after co-adding neighbouring extractions according to the colours indicated. Each extraction was first brought to rest-frame and broadened to match the central velocity dispersion. The six spectra have a S/N that varies from ($\sim150-100${\AA$^{-1}$} at 5000~{\AA} to $35-10$~{\AA$^{-1}$} at 3300~{\AA}) and central bandpasses for relevant absorption features (see table~\ref{tab:indices}) are indicated as grey shaded regions. \textit{Right:} $r$-band NGC\,3115 cutout from the DECaLS survey \citep[][]{DeyEtAl2019}. The spatial extent of each co-added annular extraction is shown, aligned with the major axis of the galaxy. Colours correspond to the final spectra shown in the left panel.}
       \label{fig:kin_extractions}
    \end{figure*}

    The kinematic fitting results are shown in the top panels of Fig.~\ref{fig:kin_extractions}, being the error bars obtained by bootstrapping the residuals from the first kinematic fit. Our results show good agreement with those from \citepalias[][]{GuerouEtAl2016}, also shown in the figure in grey. There is only a slight discrepancy for v\textsubscript{r} and $\sigma_*$ towards the outer regions, where S/N is the lowest.
    
    We performed a cubic spline fit to the v\textsubscript{r} and $\sigma_*$ points and corrected each spectrum based on the interpolation. Given that this object is clearly symmetric in its kinematic structure, we stacked the final corrected spectra in five annular extractions to maximise the S/N. These are shown in the bottom panels of Fig.~\ref{fig:kin_extractions} and are the six extractions that will be used for the following stellar population analysis.
    
\section{Stellar populations analysis}
\label{sec:stelpops}

    In this section, we describe the method used to estimate the young population fractions in NGC\,3115. We follow the one described in \citet{Salvador-RusinolEtAl2021} by performing line-strength fitting of optical and U-band indices simultaneously, using a simplified star-formation history (SFH) parametrisation. 
        
    \subsection{Stellar population models}
    \label{subsec:models}
    
    For the stellar population analysis, we again employ the E-MILES models \citep[][]{VazdekisEtAl2016}. These models, which are based on fully empirical stars, span a large wavelength range ($1680 - 50 000$~{\AA}) with a wide coverage in ages and metallicities. Because young stellar populations vary so steeply with age, we included the E-MILES extension by \citet{AsadEtAl2017} down to 6.3~{Myr} based on the PADOVA isochrones \citep[][]{GirardiEtAl2000}. For consistency, we use E-MILES models with the same set of isochrones, which span the metallicity range $-1.71$~{dex} to $+0.22$~{dex} and the age range 63~{Myr} to 17.8~{Gyr}. The initial-mass function (IMF) in massive ETGs has been shown to be bottom-heavy, particularly in their innermost regions \citep[e.g.][]{LaBarberaEtAl2013, SpinielloEtAl2014, Martin-NavarroEtAl2015b, LaBarberaEtAl2019}. To take this into account, we adopt a low-mass tapered bimodal IMF parametrised by the logarithmic slope ($\Gamma_\text{b}$) for stellar masses above $0.6$~{${\rm M}_{\odot}$}. Models are available with $\Gamma_\text{b}$ ranging from 0.3 to 3.5.

    \begin{table*}[h]
    \centering
    \caption{Index definitions used in the stellar population properties measurements}
    \label{tab:indices}
    \begin{tabular}{ccccc}
    \hline \hline
    Index name         & Blue bandpass ($\AA$) & Central bandpass ($\AA$) & Red bandpass ($\AA$) & Reference \\
    \hline
    NH3375             & 3342.000–3352.000     & 3350.000–3400.000        & 3415.000–3435.000    & \citet{ServenEtAl2011} \\
    Ni3520             & 3499.000–3508.000     & 3511.000–3530.000        & 3532.000–3548.000    & \citet{Gregg1994} \\
    Fe3619             & 3594.000-3602.000     & 3602.000-3623.500        & 3623.500-3628.500    & \citet{Gregg1994} \\
    Fe3631             & 3625.500–3628.500     & 3628.500–3637.500        & 3659.000–3674.000    & \citet{Gregg1994} \\
    Fe3646             & 3625.500–3628.500     & 3642.500–3659.000        & 3659.000–3674.000    & \citet{Gregg1994}        \\
    Fe3683             & 3659.000-3674.000     & 3676.000-3691.500        & 3693.500-3700.000    & \citet{Gregg1994}        \\
    Mg4780             & 4738.900-4757.300     & 4760.800-4798.800        & 4819.800-4835.500    &     \citet{ServenEtAl2005} \\
    H$\beta_\text{o}$  & 4821.175–4838.404     & 4839.275–4877.097        & 4897.445–4915.845    &     \citet{CervantesVazdekis2009} \\
    Mg\textsubscript{b}& 5142.625–5161.375     & 5160.125–5192.625        & 5191.375–5206.375    &     \citet{WortheyEtAl1994} \\
    Fe5270             & 5233.150–5248.150     & 5245.650–5285.650        & 5285.650–5318.150    &     \citet{WortheyEtAl1994} \\
    Fe5335             & 5304.625–5315.875     & 5312.125–5352.125        & 5353.375–5363.375    &     \citet{WortheyEtAl1994} \\
    \hline
    \end{tabular}
    \end{table*}

    \subsection{Line-strength fitting}

    Full spectral fitting, while extensively applied in optical wavelengths, still lacks a proper assessment of its applicability to the UV regime. Line-strength fitting is instead already relatively well established in this regime \citep[][]{BonattoEtAl1999, MarastonEtAl2009, LeCrasEtAl2016, Salvador-RusinolEtAl2020, DitraniEtAl2023}.
    
    While in wavelengths redder than 4000~{\AA} the main effect from $\alpha$-enhancement resides in specific absorption lines (e.g. Mg$_\text{b}$), at wavelengths bluer than 4000~{\AA} the continuum is severely affected by iron opacity. Due to this, $\alpha$-enhancement can mimic the presence of a younger component by decreasing the Fe content \citep[e.g.][]{VazdekisEtAl2015}. \citetalias{Salvador-RusinolEtAl2021}, using the MILES models with varying [$\alpha$/Fe] \citep[][]{VazdekisEtAl2015}, identified two indices that are less sensitive to the change in the abundance ratios while maximising the sensitivity to the presence of young stars, namely Fe3646 and Fe3631. This is one of the strengths of the line-strength fitting approach: selecting the specific regions where observables are maximised while decreasing the intrinsic degeneracies.

    To estimate the best-fitting parameters for our data, we use the Python package {\sc ultranest} \citep[][]{Buchner2016, Buchner2019, Buchner2021}, an open-source nested-sampling Python library already extensively used in diverse research fields. \citep[e.g.][]{DitraniEtAl2025, StefanovEtAl2025}. The main advantage of nested sampling over more traditional approaches like the Markov Chain Monte Carlo method is its increased computational speed. The code samples the likelihood uniformly within a given bound region of the likelihood $\mathcal{L}>k$ and successively increases $k$. It also provides fully sampled posteriors, allows the exploration of multimodal posterior distributions of arbitrary complexity, and retrieves the Bayesian evidence, a particular advantage when it comes to other retrieval techniques or codes which only provide the best solution, not necessarily the most representative of the distribution of the better solutions.
    
    Therefore, for each of the six spectra extracted, we compared the measured stellar population indices in the data and the models using the likelihood parametrisation and $\chi^2$ statistics
    \begin{equation}
        \ln(\mathcal{L}) \propto -\frac{\chi^2}{2} \qquad ; \qquad \chi^2 = \sum_i\left(\frac{I_{\text{data}_i}-I_{\text{model}_i}}{\sigma_i}\right)^2
        \label{eq:chi2}
    \end{equation}
    where $I_{\text{data}_i}$ and $I_{\text{model}_i}$ represent the stellar population index measurements of the $i$-th index for the observed spectrum and the models, respectively, and $\sigma_i$ is the derived uncertainty in the measurement. With this likelihood function, the code is able to get the posterior distribution for the desired parameters. Lastly, the 16th, 50th and 84th percentiles are computed from the posterior distributions. 
    
    We assumed uniform flat priors for all the parameters fitted. The index definitions used in our work can be found in Table~\ref{tab:indices}, which follow the ones selected in \citetalias{Salvador-RusinolEtAl2021}.

    \subsection{Constraining radial IMF variation in NGC\,3115}

    Before deriving what the young mass fraction is, a better understanding of the bulk stellar populations' properties in each radial bin is needed. In the wavelength range of our data, this means restricting the analysis to the reddest possible wavelengths, where the old metal-rich population dominates. Because of the shape of the spectral energy distribution (SED) of ETGs, they are much brighter in the optical range in comparison to the U-band. In addition, the throughput of the spectrograph is higher in the optical range. Therefore, because the observations were designed for good S/N in the U-band, by default, one gets `for free' an even higher quality optical spectrum. This allows us to provide constraints on the IMF and [$\alpha$/Fe] in this galaxy.

    We first perform a fit using only the reddest possible indices, including a traditional age indicator \citep[H$\beta_\text{o}$,][]{CervantesVazdekis2009}, a total metallicity indicator \citep[\text{[MgFe]'}\footnote{[MgFe]'=$\sqrt{\text{Mg}_\text{b}\left(0.72\,\text{Fe5270}+0.28\,\text{Fe5335}\right)}$},][]{ThomasEtAl2003}, and an IMF indicator \citep[Mg4780,][]{ServenEtAl2005}. We then modelled the stellar population as a single SSP and used the line-strength fitting to find the best combination of stellar population parameters that best match these lines. By default, the method provides light-weighted solutions.

    \begin{figure}[h]
       \centering
       \includegraphics[width=\columnwidth]{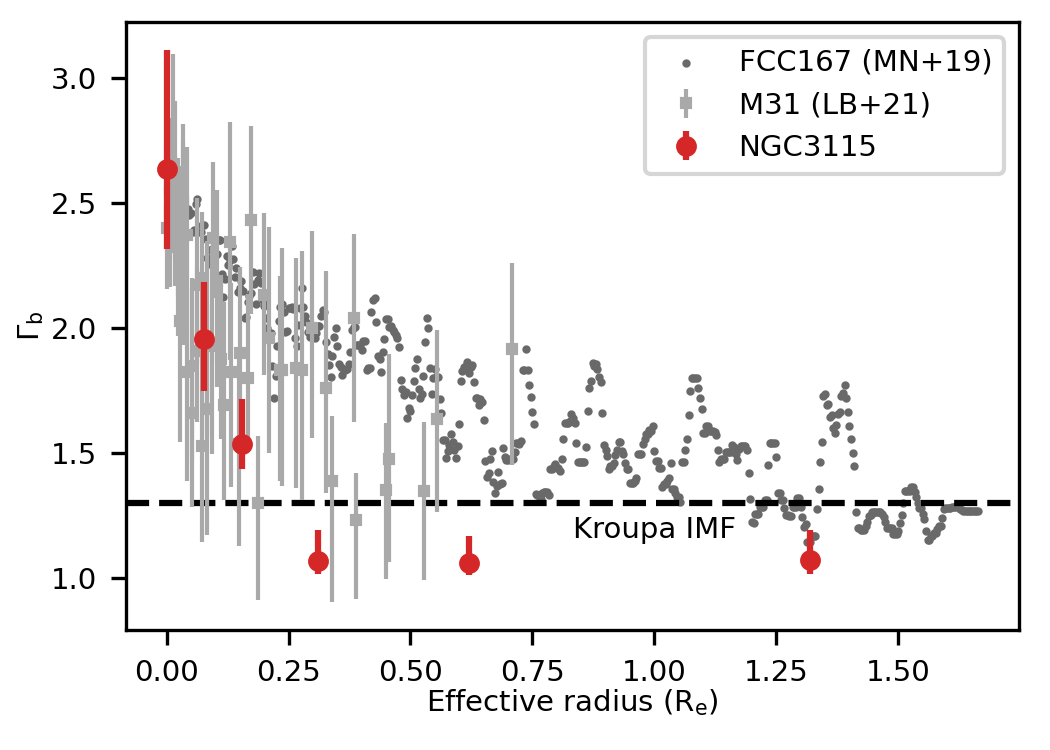}
       \caption{IMF variation parametrized by $\Gamma_\text{b}$ in NGC\,3115 within 2~{R\textsubscript{e}}. Two IMF measurements are shown for comparison: the massive lenticular FCC\,167 \citep[][]{Martin-NavarroEtAl2019a} and the massive spiral M\,31\citep[][]{LaBarberaEtAl2021}. A clear negative gradient is seen for all three galaxies, going from a very bottom-heavy IMF in the inner region to a Kroupa-like IMF in the outer bins.}
       \label{fig:IMF}
    \end{figure}

    Figure~\ref{fig:IMF} shows a clear gradient in $\Gamma_b$ for NGC~3115, behaviour similar to what has been found for other massive ETGs \citep[e.g.][]{LaBarberaEtAl2019}. We thus overplotted the results for FCC\,167, another massive lenticular galaxy, in this case in the Fornax cluster \citep[M$_*\sim 9.85 \times10^{10}$~{M$_\odot$},][]{Martin-NavarroEtAl2019a}, and for M\,31, a massive spiral that also hosts a two-component (bulge$+$disc) structure. As can be seen in Fig.~\ref{fig:IMF}, the overall behaviour measured in NGC\,3115 is also similar to these other galaxies: high $\Gamma_\text{b}$ values towards the centre and approaching a Kroupa-like IMF in the outer regions. Moreover, the departure from a standard IMF indeed happens within the pressure-dominated component of NGC\,3115, where one would expect properties to be the most similar to those of other massive ETGs.

    A caveat in this analysis is that Mg4780 tends to disagree with other IMF indicators with increasing galactocentric distance \citep{LaBarberaEtAl2016, Martin-NavarroEtAl2015b}. However, given that the strongest variations are seen within the innermost regions of NGC\,3115, predictions should be safe in this regime. Unfortunately, this is the only IMF indicator within the wavelength range probed by our observations; thus, a more detailed assessment of the IMF properties within NGC\,3115 is beyond the scope of this work. For more details on the impact of the IMF on the young-star mass fraction derivation, see Appendix~\ref{app:imf_comp}, where we perform the line-strength fitting using a fixed $\Gamma_\text{b} = 1.30$, essentially a Kroupa IMF \citep[][]{Kroupa2001}, at all radii.

    \subsection{SFH parametrization}
    \label{subsec:sfh_par}
    
\begin{figure}[h]
       \centering
       \includegraphics[width=\columnwidth]{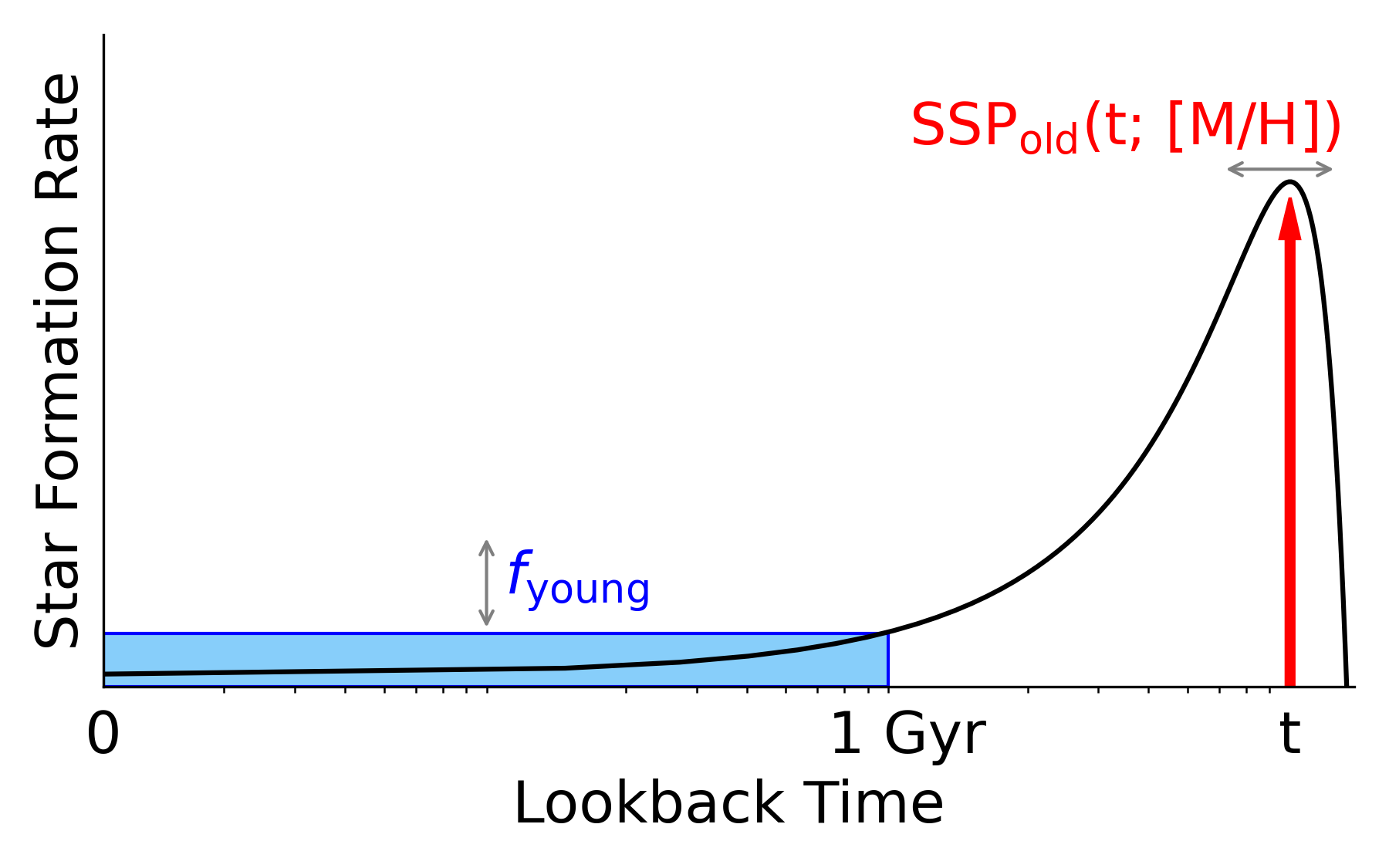}
       \caption{Illustration exemplifying the SFH model used for the line-strength fitting. The bulk of the stellar population is modelled using a single SSP governed by an age and metallicity, while the contribution of the young stars is modelled by a constant star formation over the last 1 Gyr and varying the relative mass fraction for this component.}
       \label{fig:model}
    \end{figure}

    Following the parametrisation in \citetalias{Salvador-RusinolEtAl2021}, we model the SFH by approximating a delayed-$\tau$ model as the combination of two components: the bulk stellar population component represented by a single SSP given by an age and a metallicity, overlaid with a box model that assumes a constant star-formation rate in the last Gyr (see Fig.~\ref{fig:model}). Since these measurements are being performed in the U-band, in principle, this analysis is subject to the burst age--burst strength degeneracy \citep[][]{LeonardiRose1996}. However, since we are interested in the total contribution from all stellar populations younger than $1$~{Gyr}, this degeneracy is no longer relevant.
    
    We tried fitting a single burst (i.e. a single SSP) instead of the box model for the young component. As previously mentioned, because we are limited by the U-band, the mass fraction and the age are degenerate. For example, when trying to carry out this fit, the central extraction returned $80^{+320}_{-50}$~{Myr} for the age and $0.33^{+0.84}_{-0.18}$~{\%} mass fractions for the young component. In order to be sensitive to both parameters at the same time, one needs to access the NUV, where the slope of the underlying young population can be assessed \citep[][]{Salvador-RusinolEtAl2020}.

    Another aspect of the young component that U-band observations do not allow us to probe is the metallicity of the underlying component. Young stellar population models have metallicity as only a second parameter dependency \citep[][]{AsadEtAl2017}. The extension of the E-MILES models to younger ages is actually restricted by the stellar libraries. Considering that the best-covered models by the libraries are the predictions for $[\text{M}/\text{H}] = +0.0$~{dex}, only these models are used in the construction of the young component. In order to be sensitive to the metallicities of these young stars, one would have to go to the NUV/FUV, where their light dominates, and metal lines produced by this component could be accessed \citep[][]{BrownEtAl1997}.

    \subsection{Independent estimate using [Mg/Fe]}
    \label{subsec:mgfe}
    
    Beyond following the method from \citetalias{Salvador-RusinolEtAl2021}, we also harnessed the [Mg/Fe] information in the optical to model the U-band indices. One of the key problems in doing so is the increased opacity in stellar atmospheres at wavelengths $<4000$~{\AA} due to Fe \citep[][]{VazdekisEtAl2015}, which is the reason why \citetalias{Salvador-RusinolEtAl2021} selected indices that changed the least when [$\alpha$/Fe] was varied.
    
    An alternative approach is to select `true' Fe indices in the U-band, i.e. that strongly vary only with [Fe/H]. This means that, by keeping [Fe/H] constant and varying [Mg/Fe], they should not produce any variations in the selected true iron indices. By getting a solid estimate of [Fe/H] from the widely used Fe-sensitive spectral regions in the optical \citep[e.g. Fe4383, Fe5270, Fe5335,][]{WortheyEtAl1994} where S/N is also higher, we can check whether the U-band indices align with the [Fe/H] retrieved from the optical or if there is still a mismatch. In the latter case, the young contribution can be overlaid, and those U-band mismatches can be used in the true iron indices to constrain their presence. Varying [Mg/Fe] while fixing [Fe/H] does not result in a significant change in the continuum in the U-band, as Fe is the main opacity carrier \citep[][]{VazdekisEtAl2015}. Therefore, a different abundance pattern when it comes to the Mg content cannot mimic the presence of young stars.

    To estimate [Fe/H], [Mg/Fe] and [M/H], we used the proxy method described in \citet{VazdekisEtAl2015}. By measuring a metallicity from a Mg-sensitive ([M/H]$_\text{Mg}$) and Fe-sensitive features ([M/H]$_\text{Fe}$), [Mg/Fe] estimate is obtained from:
    \begin{equation}
    \label{eq:mgfe}
    \begin{aligned}
        [Z_\text{Mg}/Z_\text{Fe}] & = [\text{M}/\text{H}]_\text{Mg} - [\text{M}/\text{H}]_\text{Fe} \quad\\
    \end{aligned}
    \end{equation}
    and then convert it to [Mg/Fe] with: 
    \begin{equation}
    \label{eq:mgfe2}
    \begin{aligned}
        \text{[Mg/Fe]} & = 0.05 + 0.62 \times [Z_\text{Mg}/Z_\text{Fe}].
    \end{aligned}
    \end{equation}
    
    Assuming $[\text{M}/\text{H}]_\text{Fe} \approx \text{[Fe/H]}$ and using the relation from \citet{KnowlesEtAl2023}, relating [Fe/H], [Mg/Fe] and [M/H], the total metallicity can also be recovered. Lastly, overlaying the young contribution on top of the bulk component (as in Fig.~\ref{fig:model}), the line-strength fitting technique is used, this time also retrieving the [Mg/Fe] value.

    \begin{figure*}[h]
       \centering
       \includegraphics[width=\textwidth]{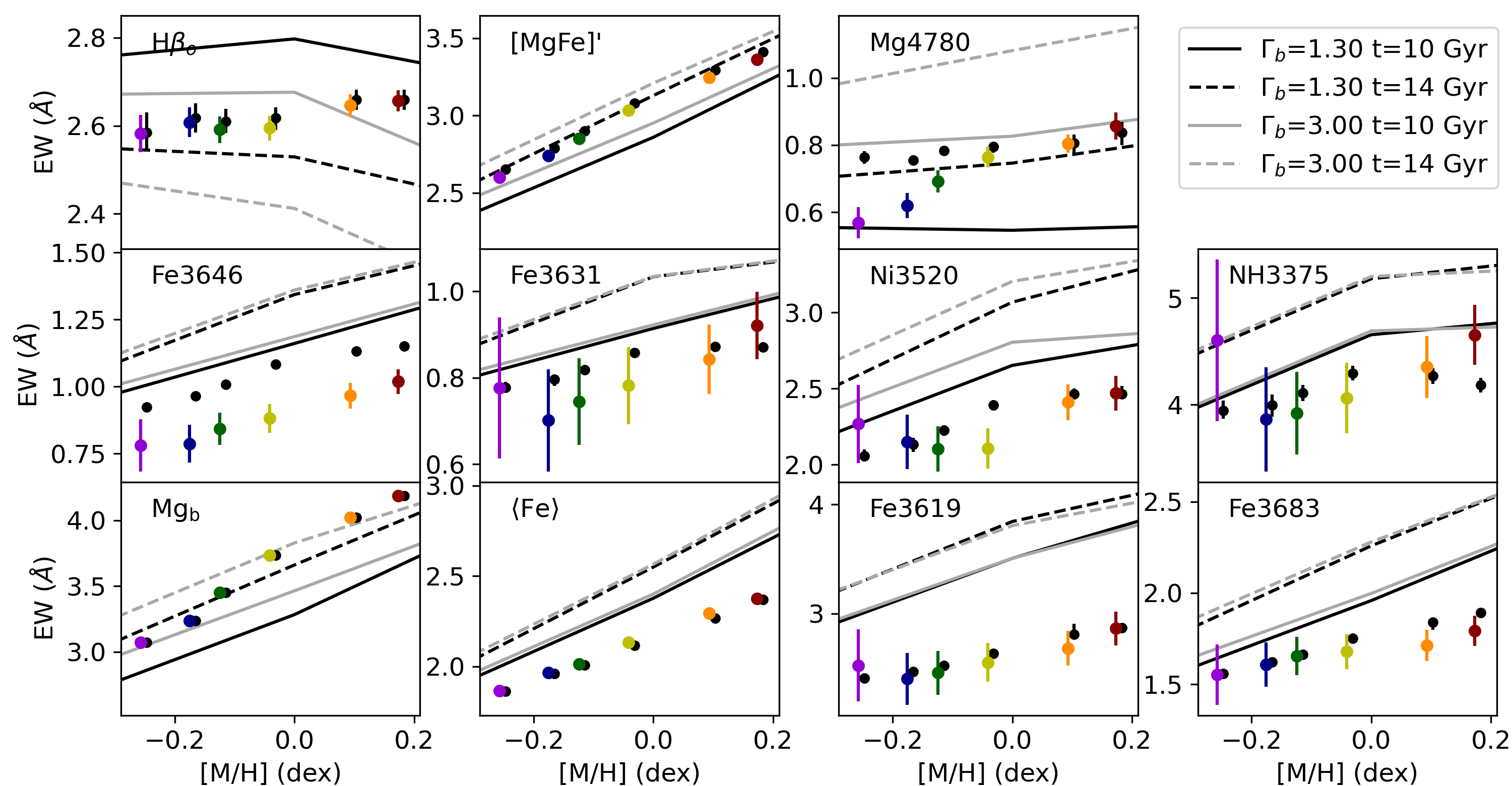}
       \caption{Index measurements for the six extractions performed in NGC\,3115 compared to model predictions. Colours are consistent with the extractions shown in Fig.~\ref{fig:kin_extractions}. The first row shows indices in the optical, the second row the U-band indices and the last row the indices used in the [Mg/Fe] retrieval method. Error bars are obtained using the equations of \citet[][]{CardielEtAl1998}. Black dots indicate the fitted models, displaced horizontally so as not to coincide with the data point loci. Lines indicate model predictions (see Sect.~\ref{subsec:models}) for varying ages, IMFs and metallicities as indicated in the labels. There is a systematic offset when it comes to U-band indices.}
       \label{fig:idx_uband}
    \end{figure*}

    As previously mentioned, the indices selected by \citetalias{Salvador-RusinolEtAl2021} are not suitable for this approach. Rather, one needs to select the indices by their response only to variations in Fe abundance. Therefore, leveraging the \citet{ConroyEtAl2018} models where individual elemental abundance can be varied, we select two other indices: Fe3619 and Fe3683 \citep[both from][, see Table~\ref{tab:indices}]{Gregg1994}. Other indices (e.g. Fe3581 or Fe3741) were discarded as variations in other elements beyond Fe (e.g. C, N, Ca, Ni) resulted in comparable changes in these indices. Using this method, beyond estimating [Mg/Fe] abundance ratio, one also gets an independent measurement of the young-mass fraction from a different set of U-band indices.

\section{Results}
\label{sec:results}

    Figure~\ref{fig:idx_uband} shows the measured indices for the six extractions colour-coded as in Fig.~\ref{fig:kin_extractions}, all matched at $\sigma = 320$~{km\,s$^{-1}$}, and, in black, the results from the fitting procedure using the parametrisation described Sect.~\ref{subsec:sfh_par}. These are compared to model predictions for different ages, metallicities and IMF shapes (as stated in the legend). The high quality of the data is evident from the small error bars \citep[obtained using equations from][]{CardielEtAl1998}. Overall, the model is successful at reproducing the line strengths.

    The optical indices H$\beta_\text{o}$ and [MgFe]', despite their small uncertainties, are properly fitted by the model, falling within the error bars. Mg4780, as previously mentioned, is a challenging IMF indicator. Despite providing consistent results in the inner regions of galaxies, it provides divergent predictions at larger galactocentric distances. This behaviour has been seen in the past compared to other IMF indicators \citep[][]{LaBarberaEtAl2016, Martin-NavarroEtAl2015b}. Therefore, it can only be properly fitted in the innermost extractions. The reason behind this offset, however, is still not known. It may be caused by varying abundance ratios that are not fully represented in the models. Another point to be considered is that IMF indicators vary strongly for bottom-heavy IMFs, but only slightly for bottom-light ones, making the predicted offset even less likely to be due to a bottom-light IMF \citep[e.g.][]{LaBarberaEtAl2016}. To mitigate the divergent behaviour of the Mg4780 effect, we take a conservative approach and fix the IMF to a standard Kroupa-like IMF for the three outer bins ($\Gamma_\text{b}=1.30$).
    
    Nonetheless, all indices in the U-band (Fe3646, Fe3631, Ni3520, NH3375) show a systematic mismatch in the values found for the models in the optical, generally showing lower values compared to model predictions. This hints at the presence of a hot component that dilutes the absorption features, which could be attributed to the presence of young stars \citep[][]{VazdekisEtAl2016}. A notable index not properly reproduced by the fitted model is Fe3646. Its strength is slightly higher for the fitted model when compared to the observations. This behaviour has already been seen in previous studies (\citetalias[e.g.][]{Salvador-RusinolEtAl2021}, \citealt[][]{Salvador-RusinolEtAl2022}). Although still unclear, the explanation for such a mismatch most likely comes from some residual effect from varying abundances (e.g. [Mg/Fe]).

    The bottom row in Fig. \ref{fig:idx_uband} displays the indices and results for the method described in Sect.~\ref{subsec:mgfe}. Varying the abundance pattern using the proxy method, we are able to fit Mg$_\text{b}$ and $\langle\text{Fe}\rangle$. We find that they are above and below model predictions, respectively, in line with the prediction of $\alpha$-enhanced stellar populations. The U-band indices Fe3619 and Fe3683 used in this fit show a strong offset with respect to the model predictions, as both the [Mg/Fe] variation and the young population presence drive the offset, reducing the index strength. Both indices, however, are also properly modelled, by taking into account the abundance pattern and the young component.

    \begin{figure*}[h]
       \centering
       \includegraphics[width=\textwidth]{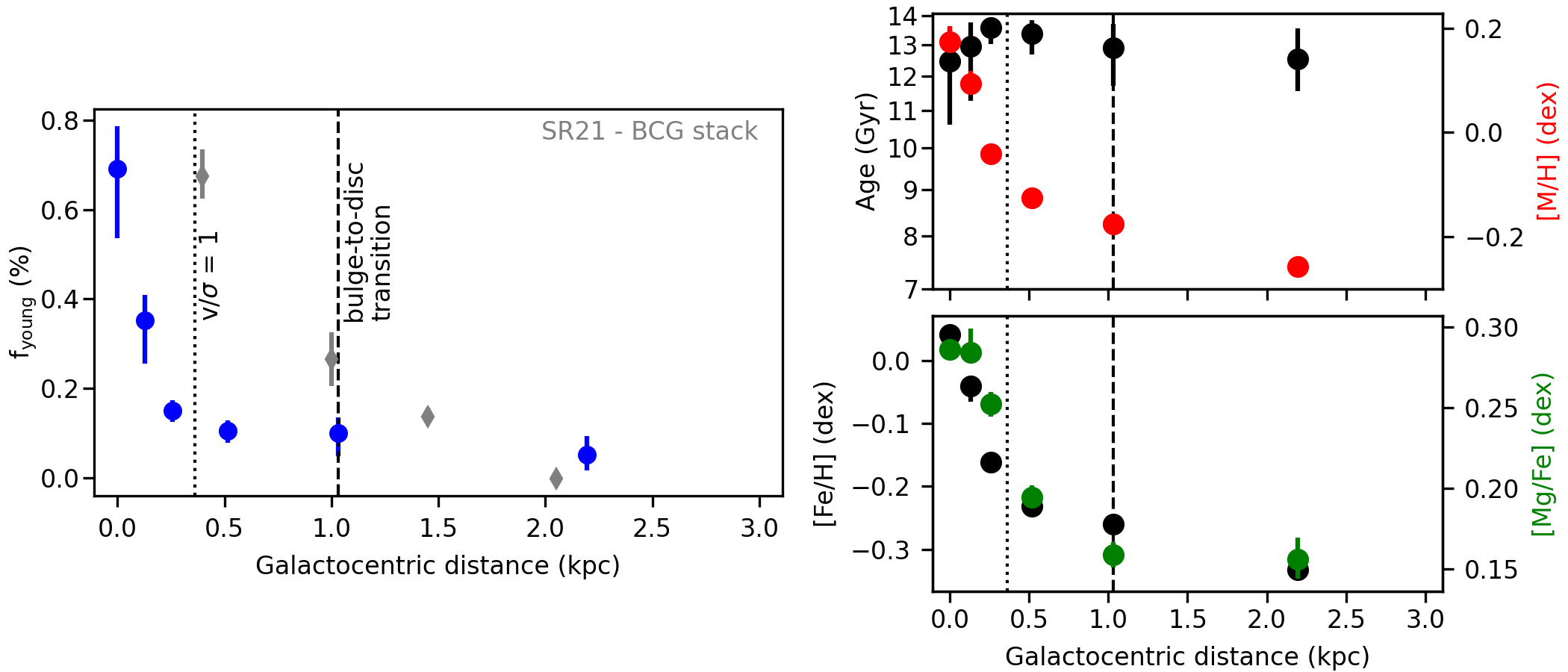}
       \caption{Results from the line-strength fitting method. \textit{Left panel:} Young stars ($<1$~{Gyr}) mass-fraction radial gradient, showing higher values towards the centre. As a comparison, \citetalias{Salvador-RusinolEtAl2021} values are overplotted, showing a similar behaviour. \textit{Right panels:} Bulk stellar population properties. From the same fit as the young star mass fraction (top panel), age and total metallicity were retrieved. The bottom panel shows [Fe/H] and [Mg/Fe] retrieved using the independent measurement and a different set of indices (see Sect.~\ref{subsec:mgfe}).}
       \label{fig:results}
    \end{figure*}

    \subsection*{Measuring the young stellar population gradient in NGC\,3115}
    
    Harnessing the mismatch in the U-band indices and the different optical indices available, we use the line-strength fitting technique to measure the key stellar population parameters to study, namely the fraction of young components (f$\rm{_{young}}$), age, [M/H], [Fe/H] and [Mg/Fe]. We derive, for each property, gradients with respect to the galactocentric distance. The left panel of Figure~\ref{fig:results} shows the results for the mass fraction of young stars, modelled by a constant star formation rate over the last 1~{Gyr}. As a reference, we indicate the transition region between the bulge and the disc (both kinematically, i.e. where $v_r/\sigma_* =1$, Fig.~\ref{fig:kin_extractions}, and photometrically \citepalias[from ][]{BuzzoEtAl2021}). We detect a negative gradient, with a mass fraction of young stars of $0.69^{+0.09}_{-0.15}$~{\%} in the centre, dropping to essentially zero at $r>1$~{kpc}. This steep decrease is in agreement with the results of \citetalias{Salvador-RusinolEtAl2021} (grey points in the figure). Although the trend is the same, the exact values and spatial extent differ slightly as the objects are different in nature. As discussed, NGC\,3115 presents at least two morphological components (bulge+disc), with the bulge dominating within 1~{kpc}; in the case of \citetalias{Salvador-RusinolEtAl2021} it is a stack of six BCGs of elliptical morphologies and high velocity dispersions ($\sigma_*>300$~{km$\cdot$s$^{-1}$}). Therefore, we expect only the inner kpc of NGC\,3115, i.e. the pressure-supported bulge-dominated region, to be comparable to the work of \citetalias{Salvador-RusinolEtAl2021}.

    To quantify the statistical significance of this detection, we use the Bayes factor \citep{MoreyEtAl2016}, which is typically represented as $\Delta \ln Z$. This allows us to compare the evidence given by the data for the presence of a young component. We thus also perform a fit with the same set of indices, but this time removing the young contribution. By comparing the Bayesian evidence $Z$ between these two realisations, we can determine which model is more favoured by the data. For reference, a model comparison with $\Delta \ln Z=3$ means one model is $e^3\approx20$ times more likely than the reference model. Comparing the two fitting schemes, we estimate the Bayesian evidence for these detections, finding $\Delta \ln Z = [4.7, 4.2, 2.7, 2.2, <1.0, <1.0]$ for each radial bin (centre towards outwards). This exercise reinforces the detection of this young component.

    Regarding the bulk of the stellar population properties retrieved, we detect a clear metallicity gradient, where the innermost parts of the galaxy host metal-rich stars ([M/H]$\sim$$0.17^{+0.03}_{-0.01}$~{dex}) decreasing to [M/H]$\sim$$-0.24^{+0.02}_{-0.02}$~{dex} in the outer bin, as seen in the top-right panel of Fig.~\ref{fig:results}, red points. In the same panel, in black points, one can see that the age gradient is instead essentially flat, with all extractions having $>12$~{Gyr}. We note that while \citetalias{GuerouEtAl2016} found a mild gradient from 13.5 to 10~{Gyr}, \citetalias{BuzzoEtAl2021} found a flat age gradient of $\sim$$11.5$~{Gyr} similar to ours. Moreover, the negative metallicity gradient is also similar among studies. While the super-solar metallicity in the centre is a consensus, the disc metallicity differs slightly in those previous studies and this one: \citetalias{BuzzoEtAl2021} found that the disc has $[\text{M}/\text{H}]$$\sim$$0.0$~{dex} and \citetalias{GuerouEtAl2016} found a lower metallicity of around $\sim$$-0.15$~{dex}, while we find an even lower metallicity of $\sim$$-0.2$~{dex}. These subtle variations reflect the diverse observational techniques and analytical approaches used in each study, along with their respective caveats. Crucially, however, the fundamental picture remains unaltered: this is an old, metal-rich galaxy exhibiting a clear negative metallicity gradient.

    Exploring the chemical composition of NGC\,3115 further, the lower right panel of Fig.~\ref{fig:results} presents the [Fe/H] and [Mg/Fe] abundances. The [Fe/H] gradient (black dots) mirrors the declining trend of [M/H], peaking in the centre at $+0.04^{+0.01}_{-0.01}$~dex and dropping to $-0.33^{+0.02}_{-0.02}$~dex. Similarly, the [Mg/Fe] ratio (green points) decreases outwards from $+0.29^{+0.01}_{-0.01}$~dex to $+0.16^{+0.02}_{-0.02}$~dex. This ratio clearly traces the transition from the inner pressure-supported bulge to the outer rotation-supported disc. Because [Mg/Fe] acts as a proxy for star formation timescales via the relative yields of $\alpha$ and Fe-peak elements \citep{ThomasEtAl1999}, this gradient indicates a slightly longer formation timescale for the disc compared to the bulge, consistent with the results from \citetalias{GuerouEtAl2016}.
    
    While \citet{Martin-NavarroEtAl2018a} found generally flatter [Mg/Fe] gradients across the broader ETG population, the two-component nature of NGC\,3115 highlights the distinct formation pathways imprinted in its structure; as previously said, we expect its pressure-supported bulge to more closely resemble typical ETGs. Finally, the apparent contradiction between a flat age gradient and a declining [Mg/Fe] gradient has been noted in other ETGs \citep[e.g.,][]{McDermidEtAl2015}. This discrepancy likely stems from the subtle spectral evolution of old stellar systems, which limits our ability to accurately assess formation timescales solely from reconstructed SFHs \citep{HartmannEtAl2025}.
    
    Finally, comparing the young-mass fractions derived from the [Fe/H]- and [M/H]-constant methods in Appendix~\ref{app:imf_comp} (see Fig.~\ref{fig:method_comp}), we notice an overall agreement between them. They coincide even when using a completely different set of indices in the U-band, further supporting the evidence for the presence of young stars in this galaxy. 

    We note that we have also assessed the impact of the measured IMF on the young mass fraction predictions. It has long been proven that using a universal IMF directly impacts the recovery of ETGs' stellar populations (see e.g. \citealt{Ferre-MateuEtAl2014}), while it has also been shown that the IMF within ETG galaxies also varies \citep{LaBarberaEtAl2016, Martin-NavarroEtAl2015b}. Indeed, \citetalias{Salvador-RusinolEtAl2021} showed that a bottom-heavy IMF boosts the young mass fractions. To account for this, we have tested how our results would vary assuming $\Gamma_\text{b} =1.3$ compared with a radially variable IMF, which are shown in Appendix\,\ref{app:imf_comp}. We find that the gradient is less pronounced when using $\Gamma_\text{b} =1.3$ (also seen in as seen in \citetalias{Salvador-RusinolEtAl2021}). Despite the lower fractions found, the Bayes factor is the same for all extractions, meaning that the data still favours a young component regardless of the IMF assumed. 
    
    In summary, we detect a negative gradient of young stars ($<1$~{Gyr}) in NGC\,3115, peaking at $\sim$$0.7\%$ in the centre, showing a consistent behaviour with previous studies for other massive ETGs. Moreover, the overall picture for the bulk of the stellar population properties follows what other literature studies have found. It hosts a ubiquitous old stellar population, with a clear decreasing metallicity gradient with super-solar metallicities towards the centre. We also find that the [Mg/Fe] abundance ratio traces the different kinematical and morphological components of NGC\,3115, hinting at the different formation timescales of each component.

\section{Discussion}
\label{sec:discussion}

    One of the key features of NGC\,3115 is its lack of a UV-upturn, being classified as UV-weak in \citet[][FUV-NUV = $1.68 \pm 0.02$]{BonattoEtAl1996}. This effectively rules out significant light contributions from hot evolved stellar sources \citep[][]{YiEtAl2011}. In Appendix~\ref{app:other_uv}, we test two alternative UV-bright models: the metal-poor HB and the metal-rich EHB scenarios. Both models are disfavoured when compared to the young hypothesis, both failing to reproduce absorption strengths in the optical and reaching unphysical results. Therefore, despite the simplicity of the assumed model, young stars are still favoured to explain the observed gradient.
    
    The presence of a gradient in this young stellar component was previously suggested by \citetalias{BuzzoEtAl2021}, who detected a bluer NUV-$r$ colour towards the centre of NGC\,3115, whilst showing a flat FUV-NUV colour gradient. The authors attributed this to young hot stars in the centre of this galaxy, claiming that a recent encounter with a companion dwarf galaxy could have been the culprit behind the recent event of star formation. They detect displaced outer isophotes, consistent with a recent ($<1$~{Myr}) peri-centric passage of such a companion. In this scenario, the young stars would have formed as the perturbed gas in the outer regions was funnelled towards the central regions of the galaxy, cooling and forming stars. While the recent pericentric passage of a satellite galaxy offers a compelling explanation for the observed SF event, establishing a direct causal relationship remains non-trivial. Quantifying this interaction requires tight constraints on several key parameters: the precise orbital trajectory of the satellite, the initial mass and spatial distribution of the gas reservoir available for displacement, and the efficiency of the companion’s tidal forces in transporting this gas inward to the central regions of the host. 

    The presence of an old thin disc with ages $>10$~{Gyr} is a crucial structural constraint for any environment-based scenario explaining the presence of young stars in this galaxy, as NGC\,3115 could not have experienced any major mergers during this period. Minor mergers, on the other hand, could still slowly deposit stars in the outskirts. Despite being considered an isolated lenticular, NGC\,3115 hosts some dwarf companions \citep[][]{PociEtAl2019, DolfiEtAl2020, Canossa-GosteinskiEtAl2024}. Past studies have already shown that the inner regions of ETGs are dominated by in-situ formed stars, whereas the ex-situ stars dominate the outer regions, thus making the scenario of an accreted origin for this young component even more unlikely \citep[][]{OserEtAl2010, Navarro-GonzalezEtAl2013, ShankarEtAl2013, CooperEtAl2015}, as the deposited stars would have to experience severe radial migration to impact the inner regions.
    
    Our comparative analysis suggests this external mechanism may be unnecessary. The mass fractions of the young component reached within NGC\,3115 are remarkably similar to those found in other massive ETGs, including relic galaxies like NGC\,1277, which have evolved without significant accretion events \citep{Ferre-MateuEtAl2017, Salvador-RusinolEtAl2022}. We therefore propose that a residual young stellar component may be an intrinsic feature of massive ETGs resulting from an in-situ process fuelled internally, e.g. by recycled material from stellar mass-loss. A strong prediction from this hypothesis is that the newly formed stars should be metal-rich, as they would have been polluted with the metals synthesised by the previous generations of stars. Unfortunately, these measurements are challenging, with the need to go to even bluer wavelengths where the light contribution from the residual component dominates.
    
    This perspective of recycled material from stellar mass loss is further reinforced by complementary constraints acquired across multiple wavelength regimes. Specifically for NGC\,3115, while optical spectroscopic data reveals a more extended SFH in the disc \citepalias[][]{GuerouEtAl2016}, UV colours point towards a bluer bulge population \citepalias[][]{BuzzoEtAl2021}. This apparent contradiction needs to be explained in the context of the timescales probed by each study. While the optical probes the more secular evolution, clearly showing that the thin disc formed over longer timescales, the UV probes the hottest stellar types. This also means that, locally, the residual SF is likely not related to the secular evolution of the galaxy, most likely being affected by other processes. Therefore, this observed discrepancy can be naturally explained by an internal recycling scenario. In this framework, debris from past stellar generations can be funnelled into the inner regions to form new stars, depleting the disc. This process effectively decouples the distribution of the young stellar component from the one predicted by optical observations.
    
    As discussed in \citetalias{GuerouEtAl2016}, there is strong evidence that NGC\,3115 hosts (or has hosted) a bar as well as disc inhomogeneities (spiral and arc features \citealt[][]{EmsellemEtAl1999};\citetalias{GuerouEtAl2016, BuzzoEtAl2021}). These structures could provide the differential torques needed to to funnel gas into the central regions of the galaxy, as has already been seen in other barred galaxies, boosting central SF \citep[][]{LinEtAl2017, ChownEtAl2019, LinEtAl2020}. \citetalias{GuerouEtAl2016} connects the detected rings with the inner and outer Lindblad resonances (ILR/OLR) at 1.0 and 2.3~{kpc}, respectively. These resonances act as attractors where gas is essentially trapped, forming the observed rings. This could pose an obstacle to the gas being funnelled further into the inner regions. NGC\,3115, however, has a more complex structure inside the ILR, hosting spiral arms that connect the ILR scales ($\sim20$~{arcsec}) to the inner $3$~{arcsec} where even a nuclear disc is observed \citep[][]{Emsellem2002}. They actually suggest that the observed spiral structures could be the remnants of an inner bar, which again provide the necessary differential torques to funnel gas inwards. It has been shown in other barred galaxies that show such inhomogeneities within the ILR are able to funnel gas into the central regions, especially in the presence of an SMBH \citep[][]{AudibertEtAl2019, AudibertEtAl2021}, so similar processes might be at play in NGC\,3115. Actually, the galaxy hosts a low-luminosity AGN \citep[][]{WrobelNyland2012, MenezesEtAl2014a, AlmeidaEtAl2018}, with faint ionised gas emission confined to the inner 1~{arcsec}. Therefore, the same inward funnelling of material required to fuel the central supermassive black hole could simultaneously fuel the residual SF we detect. This UV+optical interpretation underscores how multi-wavelength data are essential to capture the full complexity of the galaxy's evolutionary history.

    A key ingredient of star formation is molecular gas, and NGC\,3115 is known for hosting small reservoirs \citep[$\text{M}_{\text{H}_2}$$\sim$$ 1.86\times10^6$~{M$_\odot$},][]{SageWelch2006}. It was suggested that H\textsubscript{2} could be consistent with the leftover material of mass loss from dying stars. Therefore, this galaxy hosts the conditions to form new stars from recycled material of the previous generation. External contributions, however, cannot be discarded, as the atomic gas \citep[$\text{M}_{\text{HI}}$$\sim$$1.14\times10^7$~{M$_\odot$},][]{SageWelch2006} could also come from external sources.

    The bottom-heavy IMF in the inner regions, however, could pose a challenge to this scenario, as more mass gets fixed in low-mass stars without shedding their outer layers to the surrounding medium. Looking at the models, we see that while an SSP with a Kroupa-like IMF ($\Gamma_\text{b}$=1.30) releases $f_\text{gas}=38$~{\%} of the initial mass into the ISM, a bottom-heavy IMF (e.g. $\Gamma_\text{b}$=2.50) releases $f_\text{gas}=6.8$~{\%}. In both cases, there is plenty of material to be converted into stars, as even in the central region, where the stellar populations are the most bottom-heavy and show the highest young mass fraction, only 10~{\%} of the returned material would need to be converted into stars. Moreover, converting the mass fraction to an instantaneous star-formation rate, we get an $\text{SFR}$$\sim$$0.11$~{M$_\odot$\,yr$^{-1}$}. Using the total gas mass (atomic and molecular), we thus estimate a star-formation efficiency (star-formation rate per unit gas mass) of $7.5\times10^{-9}$~{yr$^{-1}$}, aligned with the expectation for other ETGs with similar stellar mass \citep[][]{KokushoEtAl2017}. Therefore, while inward gas flow from the disc to the inner regions helps reconcile the differences between purely optical \citepalias[][]{GuerouEtAl2016} and UV studies \citepalias[this work and ][]{BuzzoEtAl2021}, sustaining central star formation is not dependent on it.

\begin{figure}[h]
       \centering
       \includegraphics[width=\columnwidth]{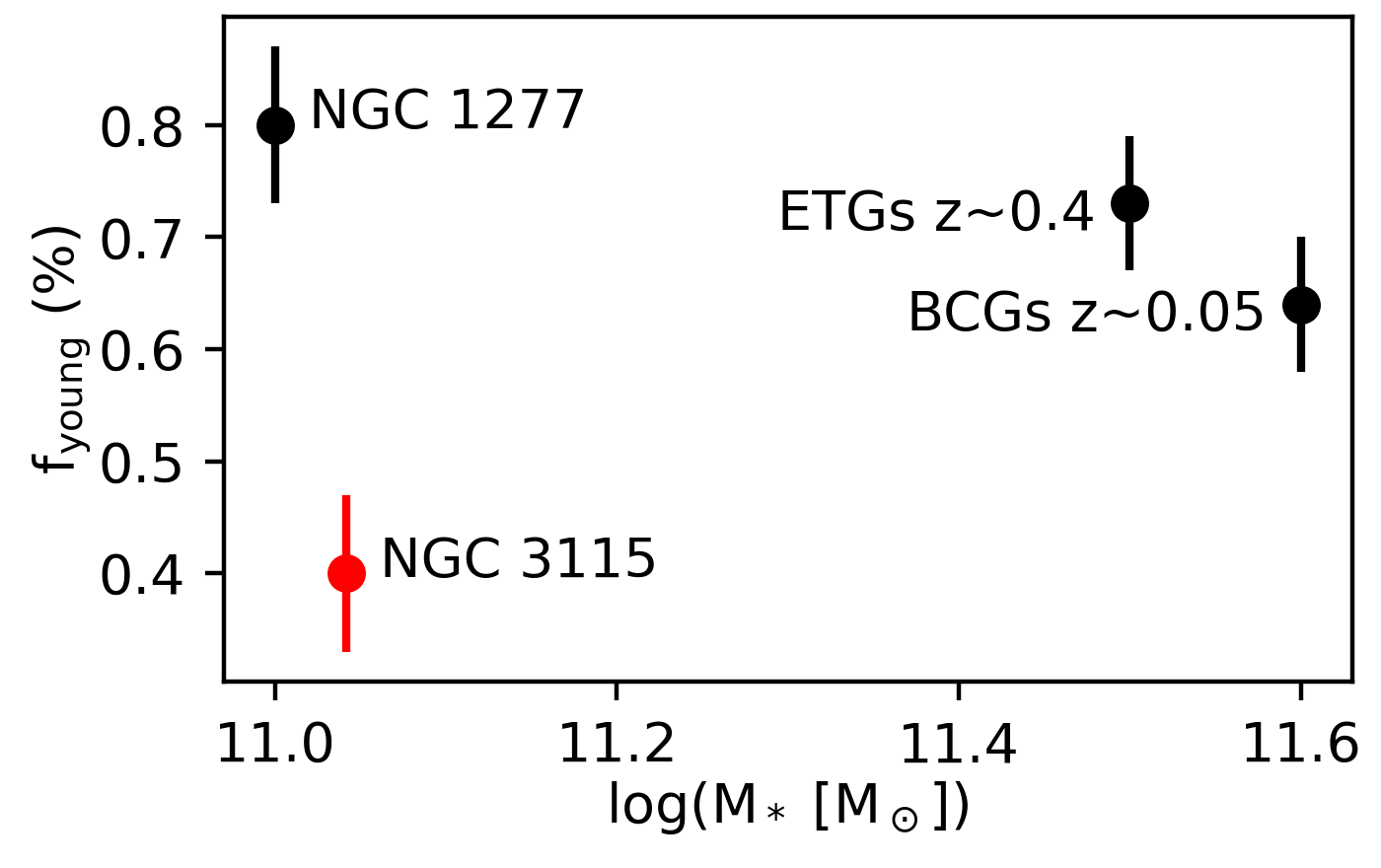}
       \caption{Young-stars ($<1$~{Gyr}) mass fraction in the inner $1$~{kpc} of NGC\,3115 compared to other massive ETGs of similar stellar masses \citep[as Fig.~{6} of ][]{Salvador-RusinolEtAl2022}. Understanding the physical processes giving rise to this component requires further exploration of the parameter space.}
       \label{fig:youngstars_etgs}
    \end{figure}

    NGC\,3115 is the first ever single ETG where a thorough analysis of the U-band indices has been conducted to derive spatially resolved information regarding the presence of a young component residual SF. Previous works have carried out similar analyses with stacks or integrated galaxy spectra (\citealt[][]{Salvador-RusinolEtAl2020,Salvador-RusinolEtAl2022}; \citetalias{Salvador-RusinolEtAl2021}). To place NGC\,3115 in this broader context, we estimate the young-star mass fraction in the inner kpc using the surface mass density form \citet{PociEtAl2019}. We present the previous results mentioned as well as the one estimated for NGC\,3115 in Fig.~\ref{fig:youngstars_etgs}, showing an evident overall sub-one-per-cent young component among the galaxies, despite their intrinsic differences. This further points towards in-situ processes dominating the fuelling of these components \citep[e.g. recycling of stellar ejecta into new generations of stars, as in][]{Salvador-RusinolEtAl2022}. While NGC\,3115 exhibits lower fractions when averaged within 1~{kpc}, current data cannot disentangle whether this stems from stochastic variations among ETGs or distinct physical drivers. Future studies are needed to fill the gaps in this diagram, exploring how mass, morphology and environment affect the presence of young components in ETGs to gain more insights into the physical drivers of such phenomena.  

    Despite the sub-one-per-cent nature of this component, cosmological simulations are particularly sensitive to it. AGN are seen as the main culprits behind shutting down star formation as well as preventing the gas from re-collapsing into new stars. The fact that this residual component appears in massive galaxies can help us assess how efficiently this interplay happens as detailed resolved studies have tried to establish this link \citep[e.g.][]{MallmannEtAl2018, doNascimentoEtAl2019, BenedettiEtAl2023, BenedettiEtAl2024}. AGN implementations, however, require ad hoc prescriptions to reproduce the properties of the present-day Universe and measuring this residual component may help us constrain the necessary physical relations in cosmological simulations. For instance, \citep[e.g.][]{IyerEtAl2025}, using the CAMELS suite of simulations, have shown that the SFH of massive galaxies is especially susceptible to variations in the AGN feedback implementation. Their late-evolution is primarily governed by how bursty the assumed nuclear activity is (what they call the $A_{\text{AGN2}}$ parameter). Therefore, populating the diagram in Fig.~\ref{fig:youngstars_etgs} might help elucidate the physics implementations within simulations and fully grasp the physical drivers of this component.

\section{Summary and conclusions}
\label{sec:conclusions}

In this paper, we have analysed the U-band long-slit spectra of NGC\,3115 gathered using the Goodman spectrograph at SOAR. By means of line-strength fitting, we derived the gradient of young stars ($<1$~{Gyr}) mass fractions for a single galaxy for the first time. The main findings of our study are summarised below:

\begin{itemize}
    \item We detect a negative mass gradient in young stellar populations in NGC\,3115, peaking at $\sim0.7$~{\%} in the centre. The computed Bayes factor $\Delta \ln Z = 4.7$ provides additional evidence for the presence of a young component and the main source producing the U-band mismatch in the line indices.
    \item This component is centrally concentrated, following the pressure-supported kinematical component and the bulge morphological component, where stellar populations are more representative of the overall ETG population.
    \item Comparing the gradients derived to \citetalias{Salvador-RusinolEtAl2021}, we find a similar behaviour despite the differences within the objects analysed, pointing to a similar trend for this component in massive ETGs.
    \item Because this galaxy does not show a UV upturn, contributions from UV-bright evolved stars are essentially discarded. Still, trial fits with different models failed to reproduce the U-band mismatch and the optical indices at the same time, further supporting the evidence of a young component in NGC\,3115. 
    \item Further supported by the NUV-$r$ gradient \citepalias[][]{BuzzoEtAl2021}, we notice that the UV provides additional information, as the gradients do not follow the SFH timescale derived in the optical \citepalias[][]{GuerouEtAl2016}. This further shows the importance of panchromatic observations to study the complete star formation histories of galaxies. 
    \item Regarding the physical drivers behind this component, our findings further support an in-situ origin for this young component. The overall agreement between the different ETGs, despite their different properties, hints at a common mode of this residual SF. The main candidate to fuel this component is the recycled material from dying stars funnelled into the inner regions of NGC\,3115, forming the young stars we detect. 
\end{itemize}

This study demonstrates that young, residual stellar components can be robustly detected and spatially mapped in single galaxies using U-band spectroscopy. Furthermore, the agreement between our results and previous studies suggests that these residual populations are a common feature of massive ETGs. While the bulk stellar properties of ETGs have been extensively studied, their late-stage mass build-up remains poorly constrained. Future panchromatic studies targeting these trace components across a wider galaxy sample will be instrumental in unravelling the mechanisms driving the secular evolution of ETGs.

\begin{acknowledgements}
    We thank the anonymous referee for the insightful comments. We thank Francesco La Barbera for providing the M\,31 IMF measurements and Ignacio Martin Navarro for the FCC\,167 IMF measurements. We also thank Michael A. Beasley for insightful discussions. Based on observations obtained at the Southern Astrophysical Research (SOAR) telescope, which is a joint project of the Ministério da Ciência, Tecnologia e Inovação do Brasil (MCTI/LNA), the US National Science Foundation's NOIRLab, the University of North Carolina at Chapel Hill (UNC), and Michigan State University (MSU). JPVB acknowledges the support of a fellowship from ”la Caixa” Foundation (ID 100010434). The fellowship code is LCF/BQ/DI23/11990084. AFM acknowledges support from RYC2021-031099-I and PID2024-162088NB-I00 of MICIN/AEI. EE acknowledges funding from the Dutch Research Council (NWO) through the award of the Vici grant VI.C.222.047. ARA acknowledges partial support from Conselho Nacional de Desenvolvimento Científico e Tecnológico (CNPq) through grants 313739/2023-4 and 402184/2026-2. 

\end{acknowledgements}

\bibliography{references_aa.bib}

\begin{appendix}

    \section{Alternative methods and prescriptions}
    \label{app:imf_comp}
    
    \subsection{IMF assumption}
    
    In order to understand the systematics in our measurement, we tested how the inferred IMF would change the results. \citetalias{Salvador-RusinolEtAl2021} have already shown that the IMF is a key ingredient in determining the mass fractions of this young stellar population. Essentially, if a bottom-heavy IMF is not taken into account, the mass fractions can drop significantly, showing a flatter gradient for the case of the BCGs analysed in that study.
    
\begin{figure}[h]
            \centering        
            \includegraphics[width=\columnwidth]{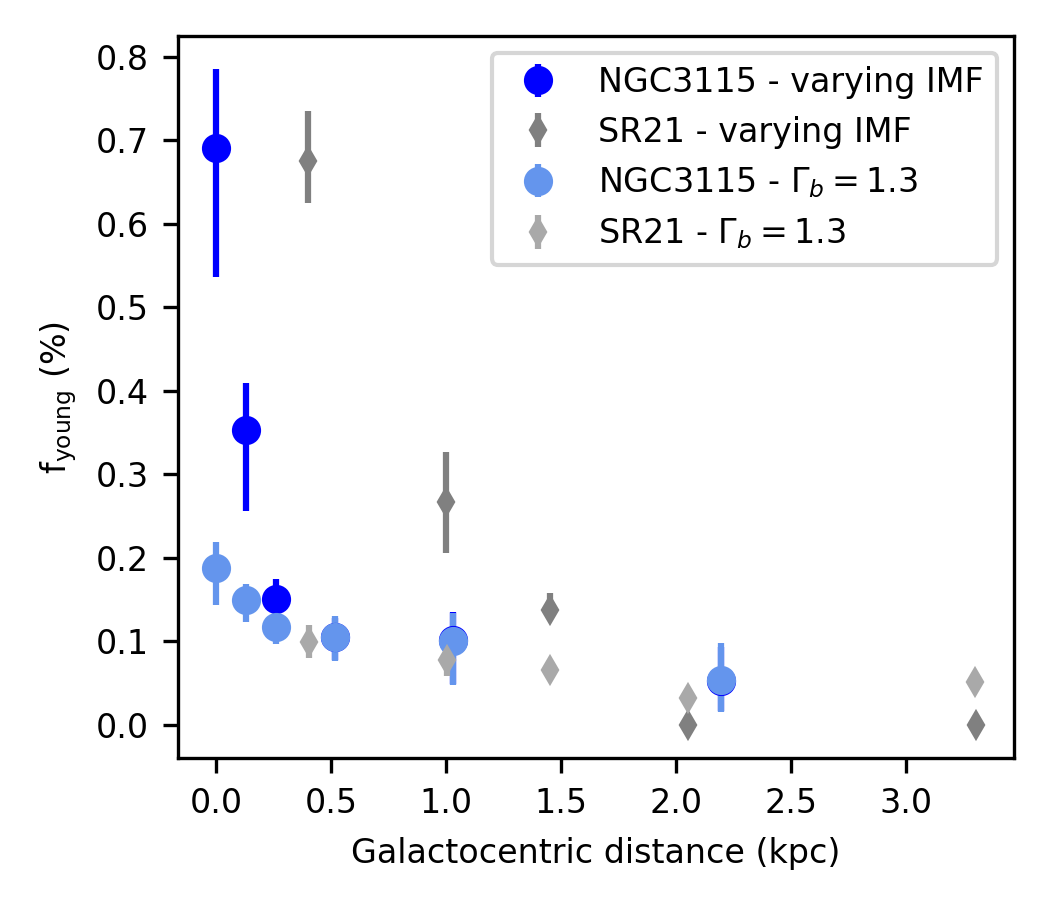}
            \caption{Impact of the IMF on the derived young mass fractions in NGC\,3115 (in blue) compared to the same effect found in \citetalias{Salvador-RusinolEtAl2021} (in grey). While the fractions are lower, the detection is also significant. The behaviour across the two datasets is similar.}
            \label{fig:imf_yfrac}
        \end{figure}

    Fig.~\ref{fig:imf_yfrac} shows the results of fixing $\Gamma_b$, removing Mg4780 (our IMF-sensitive feature) and repeating the fit. The results from varying the IMF in \citetalias{Salvador-RusinolEtAl2021} are also shown. As can be seen, the change we perceive in our data is essentially the same as the one from \citetalias{Salvador-RusinolEtAl2021}. In fact, both works follow the same trend when assuming $\Gamma_\text{b}=1.30$. An important aspect in the detection of the young component, given this alternative IMF parametrisation, is the Bayesian evidence. Despite the lower amplitude, the retrieved $\Delta \ln Z$ remain essentially unchanged when compared to the varying IMF ones. The IMF produces small changes in the spectrum shape at bluer wavelengths, being more relevant to the mass contributions of the bulk vs. the young component. Therefore, while the amplitude changes, the evidence obtained from the inclusion of the young component can remain unchanged.
    
    This goes to show that in order to provide good constraints regarding the presence of young components, one needs to harness information from the different parts of the spectrum. However, given the piling evidence for a variation in the IMF in and within ETGs, we still favour the results obtained by varying the IMF. 
    
    \subsection{[Mg/Fe] assumption}
    
\begin{figure}[h]
            \centering \includegraphics[width=\columnwidth]{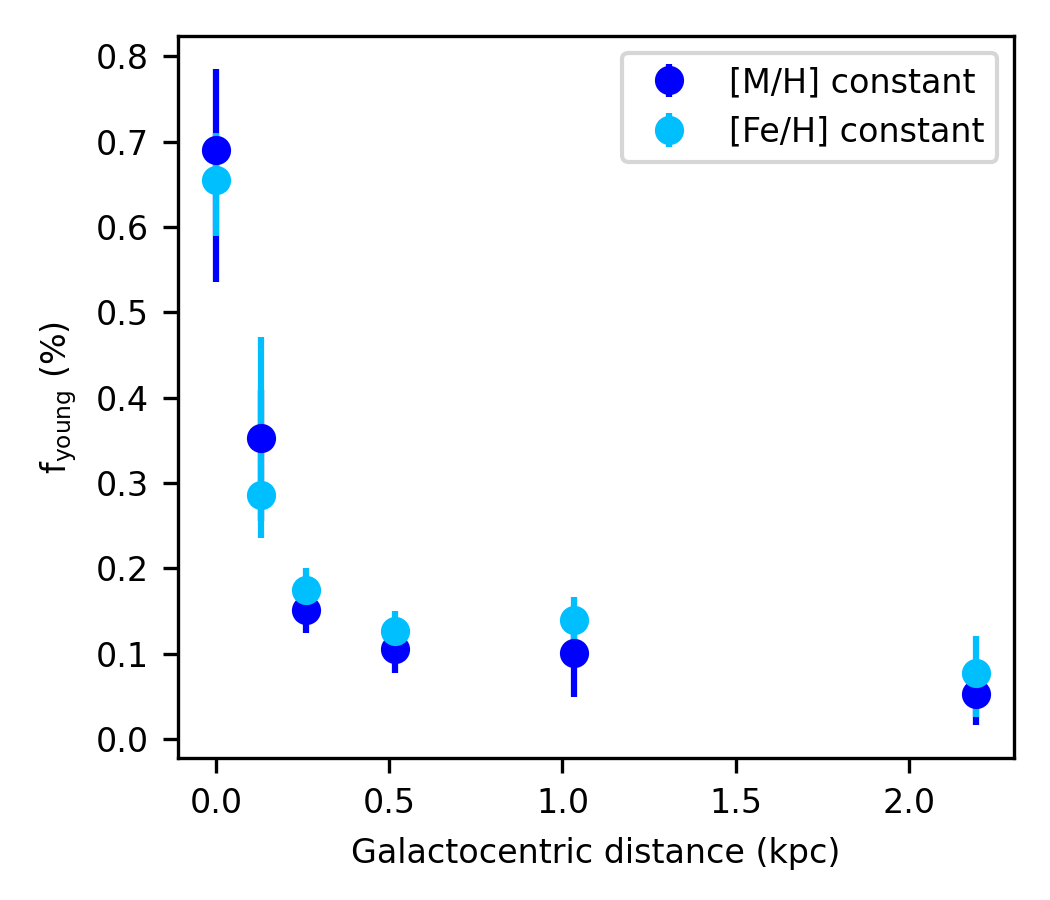}
            \caption{Young-mass fractions derived using both the standard method, avoiding U-band indices that vary with [$\alpha$/Fe] (in blue), and the [Fe/H]-constant method, where we probe the iron abundance and use U-band iron indices to estimate the young mass fractions. There is a clear agreement between the two methods within the errors, despite using a different set of indices.}
            \label{fig:method_comp}
        \end{figure}

    We also show the young fractions derived using the method described in Sect.~\ref{subsec:mgfe}, where both [Mg/Fe] and the young fractions are fitted. Since the values for the bulk stellar population are already displayed in Fig.~\ref{fig:results}, in Fig.~\ref{fig:method_comp} we show the comparison between the one obtained following the prescriptions of \citetalias{Salvador-RusinolEtAl2021} and the [Mg/Fe] method.
    
    Interestingly, despite deliberately using different U-band indices, the results obtained are strikingly similar between the two methods, showing higher values towards the centre and values agreeing within the error bars. This gives us even more confidence that the observed gradients are true, as similar results are retrieved even when different observables are used.

\section{Other UV-bright components}
\label{app:other_uv}

    One of the challenges when interpreting the UV wavelengths of ETGs is the presence of hot UV-bright components that can outshine the old metal-rich populations in this regime. The most prominent manifestation of this component is the UV upturn phenomenon, which boosts the FUV flux, reversing the expected trend of lower FUV fluxes for these systems. NGC\,3115, however, does not host a UV upturn and is actually classified as UV weak in \citet[][FUV-NUV = $1.68 \pm 0.02$]{BonattoEtAl1996}.

    However, there is a gradient with the NUV-$r$ colour \citepalias[][]{BuzzoEtAl2021}, showing an excess in its NUV emission but not in its FUV emission. This by itself already puts a tight constraint on the type of UV population present in NGC\,3115. Horizontal Branch (HB) metal-rich candidates for the UV upturn are typically characterised by heavily stripped stars which expose their inner shells, shining brightly in the UV. The same applies to post-AGB stars, which expose their cores, shedding their outer layers. This means that the contribution from these hot evolved stars is essentially excluded, as they would produce substantial FUV emission. In the scenario where the excess in UV emission is produced by younger stars, the shallower FUV emission can easily be explained by changes in the SFH. Strong FUV fluxes are produced by really young stars ($<20$~{Myr}). Considering that our young model has a longer timescale ($1$~{Gyr}), the contribution from these stars is reduced. This results in proportionally more stars in the older end of the young component, as expected for our box model and for the delayed-$\tau$ model, for example.

    Moreover, as argued in \citet{Salvador-RusinolEtAl2022}, the U-band indices are limited to constrain the presence of other UV components and only observations reaching the NUV are able to disentangle the two components. However, they point out that the overall agreement between their results (in their case NGC\,1277) and the results from the BOSS survey analysis \citep[][]{Salvador-RusinolEtAl2020}, where they had access to NUV indices, allowed them to discard other UV sources. This further supports our findings, as we also derive young-mass fractions that are consistent with these studies.

    Regardless, we also fit for other UV-bright components in the literature known for increasing the flux at bluer wavelengths: the presence of a metal-poor extended HB (EHB) and a metal-rich EHB. Using the full set of indices from before, we find a constant mass fraction of $\sim 3\%$, as H$\beta$ and [MgFe]' limit the mass contribution from the metal-poor component. With this amplitude, the U-band indices, however, are not properly fitted, as they would require a more substantial contribution from the UV-bright component. When we remove the optical indices from the fit, then the U-band indices are well-fitted, and the resulting mass fractions go from 15\% in the central extraction to 2\% in the outer ones. For the inner regions, the mass fractions are well over the predicted ones for massive ETGs \citep[$\lesssim 5\%$, e.g.][]{VazdekisEtAl1997} and result in deeper H$\beta$ than observed. Comparing the solutions obtained for the metal-poor EHB and the young scenario using the Bayes parameter, we get $\Delta \ln Z < 1.0$, meaning the metal-poor model is unfavoured to explain the U-band mismatch.

    In the metal-rich scenario, fitting for age, metallicity and mass fraction of the EHB model for the same indices, the optical indices limit the presence of this hot HB component, as it also predicts a deeper H$\beta$ absorption. Using this model and fitting for all the indices, we recover a mass fraction of $\sim12\%$. However, the U-band indices are also not well fitted. If, rather, we fix the age and metallicity to the ones found before, we then are able to fit the U-band indices, but require a mass fraction of $\sim55\%$ in the central regions, similar to what \citet{Salvador-RusinolEtAl2022} found for NGC\,1277. This, according to the authors, is a much higher fraction compared even to the most extreme examples of EHB stellar populations \citep[e.g. $\omega$Cen,][]{DCruzEtAl2000}.
    
    These exercises of trying to fit the U-band with different models for this galaxy show that the young mass fraction we measure due to the index dilution in the U-band is indeed the best-fitting model we are able to find, matching also the photometric gradients found in \citetalias{BuzzoEtAl2021}. This is not to say that this is the only contribution. Most likely, several sources coexist, but from our results, the young star contribution seems to be the most significant one. 

\end{appendix}

\end{document}